\documentclass[conference]{IEEEtran}
\IEEEoverridecommandlockouts
\usepackage{cite}
\usepackage{amsmath,amssymb,amsfonts}
\usepackage{hyperref}
\usepackage{algorithmic}
\usepackage{graphicx}
\usepackage{textcomp}
\usepackage{xcolor}
\usepackage{url}
\def\BibTeX{{\rm B\kern-.05em{\sc i\kern-.025em b}\kern-.08em
    T\kern-.1667em\lower.7ex\hbox{E}\kern-.125emX}}
\newcommand{\Nmax}{\ensuremath{N_{\max}}}

\begin{document}

\title{Closed-Loop L4S-as-a-Service in 5G-Advanced: NEF–PCF Control with NWDAF-Driven Assurance}

\author{
    \IEEEauthorblockN{\textbf{Ameer Shohail L}}
    \IEEEauthorblockA{\textit{5G CORE Design Architect}\\
    \textit{Communications Service Provider}, \\
    Riyadh, KSA  \\
    Email: ameershohail@ieee.org, \\
     ameershohail@gmail.com}
    \and  
    \IEEEauthorblockN{\textbf{Ashish Goswami}}
    \IEEEauthorblockA{
        \textit{Department of ECE}\\
        Amrita School of Engg., Bengaluru,\\
        Amrita Vishwa Vidyapeetham, India\\
        Email: goswami.ashish@ieee.org,\\
        g\_ashish@blr.amrita.edu
    }
}

\maketitle

\begin{abstract}
Ultra-low latency services in 5G-Advanced demand deterministic delay and high-fidelity congestion signaling beyond peak throughput. While the Low Latency, Low Loss, Scalable Throughput (L4S) architecture enables sub-millisecond queuing through ECN-based feedback and Dual-Queue Coupled AQM, its integration within the 5G Core (5GC) remains functionally siloed. Current 3GPP Release 18/19 specifications provide mechanisms for L4S enablement, but they do not define a unified closed-loop framework that links application intent to verified service outcomes. To address this gap, we propose Closed-Loop L4S-as-a-Service (C-L4SaaS), an architectural framework that orchestrates the Network Exposure Function (NEF), Policy Control Function (PCF), and Network Data Analytics Function (NWDAF) for automated latency assurance. The framework translates high-level intent into enforceable PCC rules and uses NWDAF-driven compliance analytics, derived from User Plane Function (UPF) measurements, to trigger bounded policy adaptations. We model this interaction as a discrete-time feedback system and derive stability conditions and signaling overhead bounds to guide parameter selection under volatile wireless conditions. The proposed core-driven orchestration provides a standards-aligned path to expose, assure, and govern managed low-latency services in 5G-Advanced ecosystems.

\end{abstract}

\begin{IEEEkeywords}

QoS\slash QoE, 5G-Advanced, L4S, Closed-loop QoS assurance, NWDAF analytics, NEF–PCF policy control, 5G Core
\end{IEEEkeywords}

\section{Introduction}
Ultra-low latency has shifted from a best-effort objective to a service requirement in 5G-Advanced, driven by interactive and time-sensitive applications that are sensitive to tail delay rather than average throughput. In parallel, the Internet congestion-control community has introduced Low Latency, Low Loss, Scalable Throughput (L4S) ~\cite{rfc9330} to provide high-frequency congestion signaling using Explicit Congestion Notification (ECN), enabling scalable throughput while holding queuing delay close to a low target. L4S ~\cite{rfc9331} is therefore attractive for mobile networks that must multiplex heterogeneous traffic while preserving low delay for selected flows.

Despite this promise, operationalizing L4S within the 5G Core (5GC) ~\cite{etsi23501r18} remains fragmented. The 5GC already contains the functional building blocks needed to expose services, enforce policy, and generate analytics. The Network Exposure Function (NEF) can expose APIs \cite{camara} to applications, the Policy Control Function (PCF) can translate service requirements into enforceable Policy and Charging Control (PCC) rules \cite{ts23503}, and the Network Data Analytics Function (NWDAF) can provide analytics to network functions. However, current practice and specifications do not define an integrated closed-loop workflow that binds application intent to measurable and verifiable service outcomes for L4S-enabled flows, especially under volatile wireless capacity conditions. This aligns with the broader zero-touch automation objective, where closed-loop assurance is treated as a first-class management capability \cite{zsm002, etsi104041}.

This paper addresses that gap by proposing Closed-Loop L4S-as-a-Service (C-L4SaaS), a standards-aligned architecture that couples service exposure, policy enforcement, and analytics-driven assurance into a single control loop, consistent with zero-touch closed-loop management principles \cite{zsm002}. Service-centric congestion-control abstractions have also been explored in the IWQoS community, motivating the ``as-a-service'' framing adopted here \cite{iwqos2024_ccaas}. C-L4SaaS enables an application to request managed low-latency treatment via a NEF-exposed API (prior work \cite{11317283})\cite{camara}, maps the request into PCC policies at the PCF, and applies L4S-oriented treatment through SMF-directed UPF configuration \cite{ts29512}. NWDAF closes the loop by deriving compliance analytics from UPF measurements, enabling the PCF to apply bounded policy adaptations when service-level objectives are not met and to iteratively reassess compliance.\newline
The contributions of this paper are as follows: First, we present an end-to-end architectural blueprint for C-L4SaaS and map it to 5G service-based functions and control points required to expose and govern L4S as a managed capability. Second, we formulate the assurance loop as a discrete-time feedback system, defining compliance metrics, an adaptation signal, and a bounded policy update mechanism. Third, we derive stability and signaling overhead bounds that guide parameter selection and scalability under time-varying wireless conditions. Collectively, these results provide a practical and standards-aligned path to expose, assure, and govern managed low-latency services in 5G-Advanced ecosystems.

\section{CLOSED-LOOP L4S-AS-A-SERVICE: ARCHITECTURE AND WORKFLOW}

\subsection{Architectural Overview and Design Rationale}


The 5G Core (5GC) Service-Based Architecture exposes, controls, and assures network capabilities through functions with clear responsibilities. C-L4SaaS uses this model to operationalize Low Latency, Low Loss, Scalable Throughput (L4S) as a managed service rather than a static configuration. Since enablement alone does not guarantee outcomes, C-L4SaaS applies a closed-loop that binds application intent to authorized policy, enforces L4S treatment on selected service data flows, and continuously verifies compliance with bounded corrective actions.

\begin{figure*}[t!]
\centering
\includegraphics[width=\textwidth]{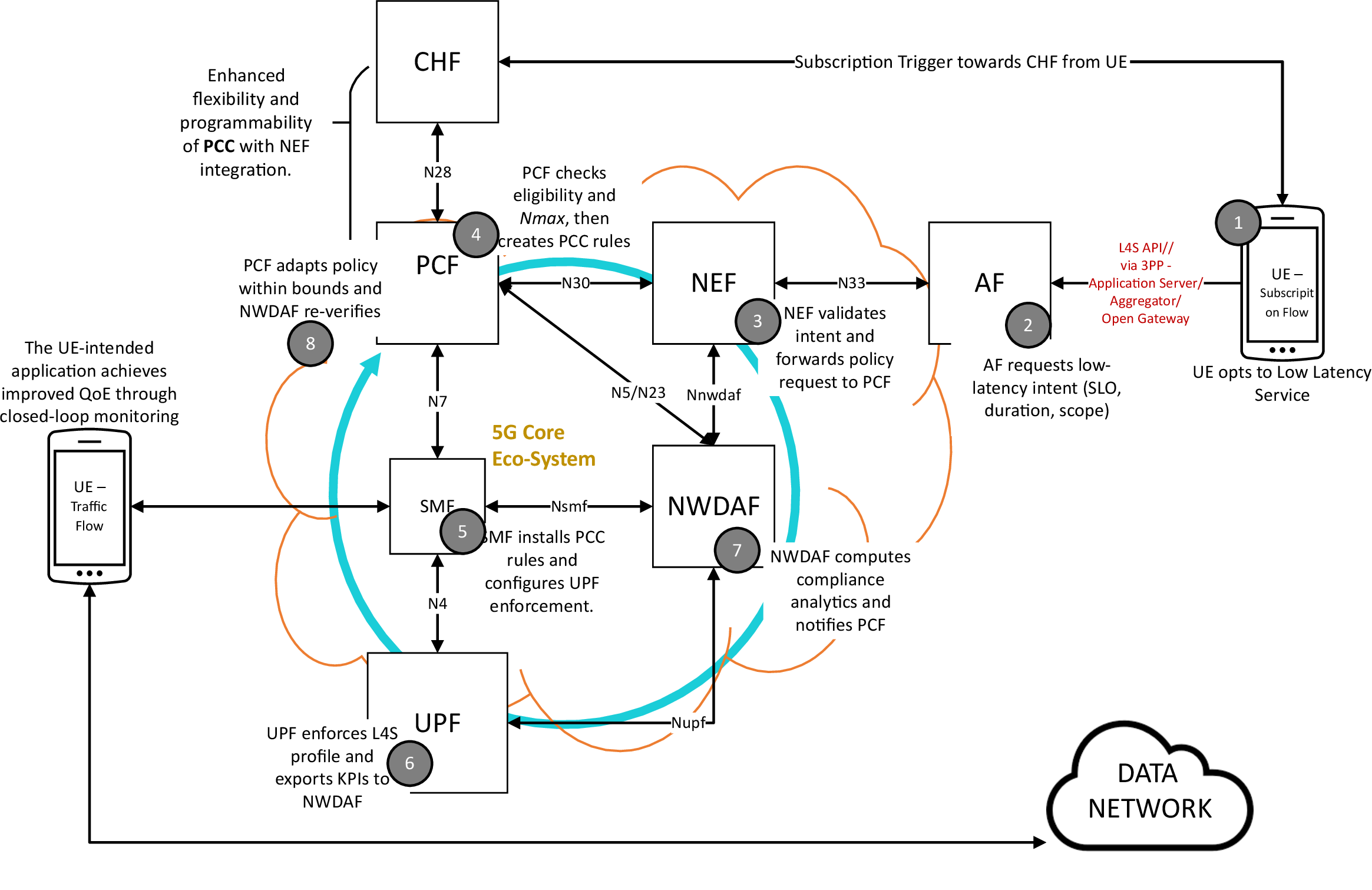}  
\caption{C-L4SaaS closed-loop architecture}
\label{fig:C-L4S_Architecture}
\end{figure*}

As shown in Fig.~\ref{fig:C-L4S_Architecture}, the framework relies on three pillars. NEF provides governed service exposure, PCF translates intent into enforceable PCC rules, and NWDAF derives compliance analytics from user-plane measurements to drive bounded adaptation. SMF and UPF realize enforcement, while NWDAF verifies SLO compliance over time.

\subsection{Functional Roles of NEF, PCF, NWDAF, SMF, and UPF}




\textit{NEF (Intent Exposure and Governance)}:NEF is the governed ingress for third-party or operator AFs to request managed low-latency service. It enforces authentication, authorization, tenant isolation, and rate limits, then forwards the request for PCF policy evaluation.

\textit{PCF (Intent Translation and Policy Actuation):} PCF is the decision point for granting the service. It translates intent into PCC rules bound to subscriber and slice context and to flow descriptors, and acts as the bounded actuator by updating a small set of policy variables within predefined limits.

\textit{SMF/UPF (Enforcement Plane):} SMF realizes PCC policies at session level and configures UPF enforcement. UPF applies the authorized handling profile per flow and exports measurements required for compliance analytics, avoiding global changes that could impact other traffic.

\textit{NWDAF (Compliance Analytics and Feedback):} NWDAF aggregates user-plane measurements over windows and produces compliance analytics. PCF consumes these analytics as the closed-loop feedback signal, with windowing used to reduce noise and transient effects.

\subsection{Primary Control Strategy: Admission-Limited Managed Low-Latency Service}

C-L4SaaS adopts admission control as the primary bounded actuator for stability and portability across multi-vendor deployments. The PCF maintains an admission limit \Nmax per slice or policy domain, while NWDAF compliance analytics indicate whether admitted load meets SLO targets. When compliance degrades, PCF tightens \Nmax to protect tail delay; when stable, it relaxes \Nmax conservatively within bounds. Secondary actions, such as bounded priority offsets or GBR adjustments, are reserved for sustained non-compliance.

A compact discrete-time actuation rule for the primary control variable is:
\begin{equation}
N_{\max}[n+1]=\Pi_{[0,N_{\mathrm{cap}}]}\left(N_{\max}[n]-K\,e[n]\right),
\end{equation}
where $e[n]$ is computed from NWDAF compliance analytics (e.g., tail-delay deviation from target), $K$ is a conservative gain, and $\Pi$ enforces bounds and prevents oscillation.

\subsection{End-to-End Workflow for Closed-Loop Operation}

\begin{figure*}[t!]
\centering
\includegraphics[width=\textwidth]{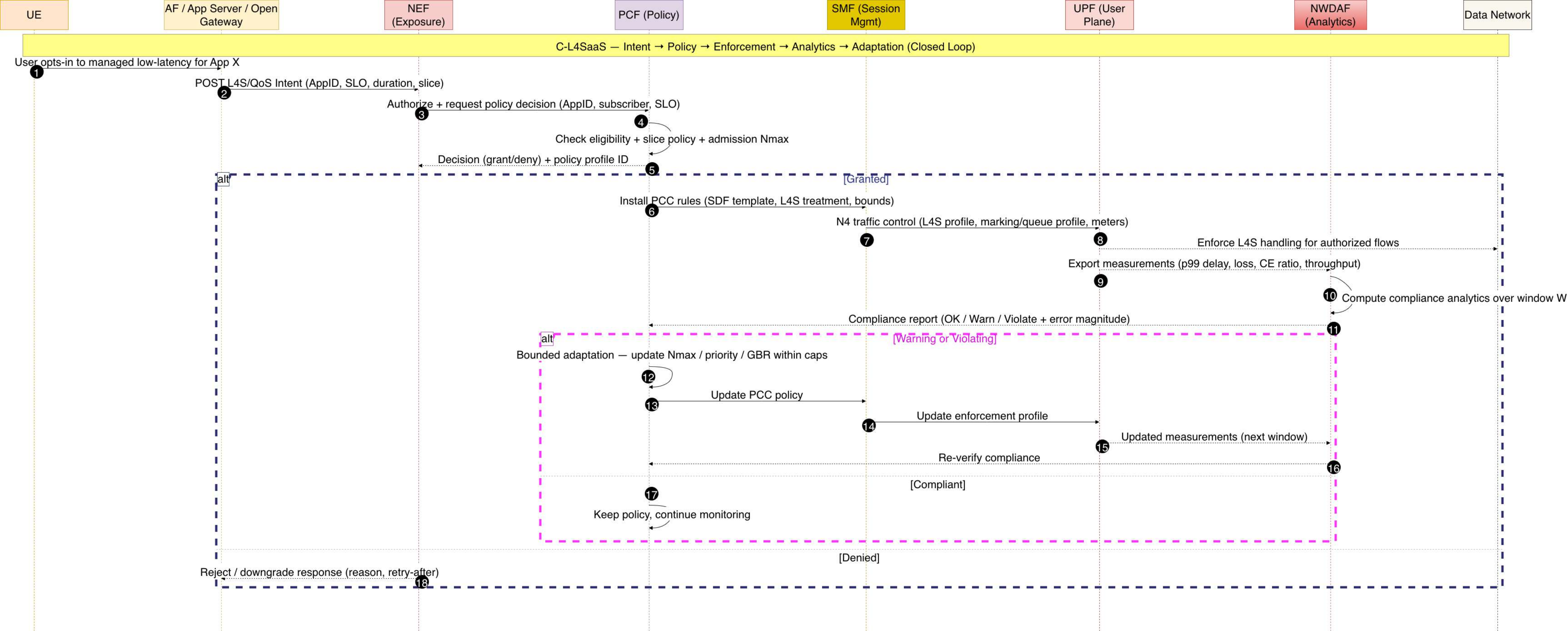}  
\caption{C-L4SaaS architecture and closed-loop assurance across AF/NEF, PCF, SMF/UPF, and NWDAF}
\label{fig:C-L4S_CallFlow}
\end{figure*}

The C-L4SaaS workflow is defined as an auditable sequence that binds each action to a responsible 5GC function, as shown as Fig.~\ref{fig:C-L4S_CallFlow}.
\begin{enumerate}
    \item \textbf{Intent request (AF $\rightarrow$ NEF).}
    An application function requests a managed low-latency service for a target flow, including SLO targets (e.g., tail delay threshold), duration, and scope (subscriber, slice, and application identifiers where applicable).
    NEF validates the request and applies governance controls.

    \item \textbf{Authorization and policy translation (NEF $\rightarrow$ PCF).}
    NEF forwards a policy request to PCF.
    PCF evaluates eligibility, checks current admission state $N_{\max}$, and determines whether the service can be granted without violating slice policy constraints.

    \item \textbf{PCC rule installation (PCF $\rightarrow$ SMF).}
    If granted, PCF generates PCC rules that bind the intent to a service data flow template and to an enforcement profile.
    The PCC rules are delivered to SMF for session realization.

    \item \textbf{Enforcement realization (SMF $\rightarrow$ UPF).}
    SMF configures UPF for the authorized flow handling profile and enables measurement export for compliance analytics.
    UPF applies the configured handling behavior to the target flow only.

    \item \textbf{Compliance analytics generation (UPF $\rightarrow$ NWDAF).}
    NWDAF collects UPF measurements over a fixed analytics window $W$ and computes compliance analytics, including tail delay, packet loss ratio, and ECN marking frequency.
    
    \item \textbf{Bounded adaptation and re-verification (NWDAF $\rightarrow$ PCF $\rightarrow$ SMF/UPF).}
    If NWDAF indicates sustained non-compliance, PCF applies bounded adaptation.
    The primary adaptation is adjustment of $N_{\max}$, complemented by bounded secondary actions when needed. NWDAF then re-verifies compliance in subsequent windows, completing the closed loop.
\end{enumerate}

\subsection{Compliance States and Bounded Actions}
To ensure predictable behavior, C-L4SaaS defines discrete compliance states derived from NWDAF analytics. Table I provides the default compliance thresholds and the associated bounded PCF actions. The default values are anchored to standardized QoS characteristics using \textbf{5QI 3} as a representative interactive low-latency class.

\begin{table}[t]
\caption{NWDAF COMPLIANCE STATES AND BOUNDED POLICY ACTIONS (C-L4SaaS CLOSED-LOOP)}
\label{tab:nwdaf-compliance}
\centering
\renewcommand{\arraystretch}{1.0}
\scriptsize
\begin{tabular}{|l|p{2.35cm}|p{1.05cm}|p{2.35cm}|}
\hline
\textbf{State} & \textbf{Condition} & \textbf{Meaning} & \textbf{Action} \\
\hline
\textbf{OK} &
$(\hat d_{99}(W) \le 50\,\mathrm{ms}) \ \textbf{and}\  (\hat{\ell}(W) \le 10^{-3})$ &
Met &
Hold \\
\hline
\textbf{Warn} &
$(50 < \hat d_{99}(W) \le 55\,\mathrm{ms}) \ \textbf{or}\ \newline (\hat{\ell}(W)\ \rightarrow\ 10^{-3})$ &
Drift &
Limit admits; $\Delta w$ \\
\hline
\textbf{Violation} &
$(\hat d_{99}(W) > 55\,\mathrm{ms}) \ \textbf{or}\  (\hat{\ell}(W) > 10^{-3}) \ \textbf{for}\ k=3\ \text{windows}$ &
Sustained &
$\ N_{\max}\!\downarrow$; prio/GBR (cap) \\
\hline
\textbf{Unstable} &
Oscillation or overhead cap &
Risk &
Freeze; revert; $W\uparrow$; hysteresis \\
\hline
\end{tabular}
\vspace{1mm}
\begin{flushleft}\scriptsize
\textit{Defaults}: $W=2\,\mathrm{s}$, $p=99$, $\delta=5\,\mathrm{ms}$, $k=3$. \textit{Anchoring}: 5QI 3 PDB $=50\,\mathrm{ms}$ and PER $=10^{-3}$ (3GPP TS 23.501 standardized QoS characteristics ~\cite{3gpp23501}). \newline
\newline \textit{Abbrev.}: OK=Compliant; Warn=Warning; prio=priority; GBR=Guaranteed Bit Rate; $W$=window; $k$=consecutive windows; $\ N_{\max}$=max. admitted flows; $\Delta w$=policy weight offset; $\hat d_{99}$=estimated 99th-percentile delay; $\hat \ell$=estimated loss.
\end{flushleft}
\end{table}

\subsection{PCF Actuation Space and Guardrails}

C-L4SaaS restricts actuation to a small, bounded set of policy knobs to ensure stability and operational safety. Table~II lists the primary and secondary control variables and their guardrails. The primary actuator is $N_{\max}$ due to its stability and portability.

\begin{table}[t]
\caption{PCF control knobs ($u$) for closed-loop L4S assurance}
\label{tab:pcf-control-knobs}
\centering
\renewcommand{\arraystretch}{1.0}
\scriptsize
\setlength{\tabcolsep}{2.4pt}
\begin{tabular}{|p{1.55cm}|p{1.55cm}|p{2.05cm}|p{2.05cm}|}
\hline
\textbf{PCF Control Knob ($u$)} & \textbf{Where Applied} & \textbf{Primary Effect} & \textbf{Bounds / Guardrails} \\
\hline
\textbf{Admission limit} $N_{\max}$ & NEF/PCF auth. & Prevent overload; protects tail delay & $0\!\le\!N_{\max}\!\le\!N_{\mathrm{cap}}$; update $\le$1/$W$ (e.g., $W\!=\!2$\,s); hysteresis \\
\hline
\textbf{Prio./weight offset} $\Delta u$ & SMF$\rightarrow$UPF (opt. RAN map) & Reduces tail delay under contention & Slice caps; bounded step size $\Delta u$ \\
\hline
\textbf{GBR within caps} & PCF via SMF QoS & Stabilizes delay under bursts & $\mathrm{GBR}_{\min}\!\le\!\mathrm{GBR}\!\le\!\mathrm{GBR}_{\max}$; only for sustained violation \\
\hline
\textbf{Enforcement profile} & UPF profile & Alters marking/queue handling aggressiveness & Pre-validated profiles; rate-limit changes (e.g., 1 per 2--3 windows) \\
\hline
\textbf{Analytics window} $W$ & NWDAF config & Improves stability under noise & $W\!\in\!\{2,4,6\}$\,s; increase $W$ during oscillation \\
\hline
\end{tabular}
\vspace{1mm}
\begin{flushleft}\scriptsize
\textit{Abbrev.}: PCF control knobs: $u$; GBR=Guaranteed Bit Rate; $W$=analytics window; $N_{\mathrm{cap}}$=admission cap; $\Delta u$=bounded policy step.
\end{flushleft}
\end{table}

\subsection{Signaling Overhead Bound for Closed-Loop Scalability}
Because closed-loop assurance consumes control-plane resources, C-L4SaaS includes an explicit overhead guardrail. If
$N$ managed sessions are subject to closed-loop updates at interval $T_s$ with average per-update message size $B$,
the control-plane rate is bounded by
\begin{equation}
R_{\mathrm{ctrl}} \le \frac{N B}{T_s}.
\end{equation}
This bound motivates the use of windowed analytics and conservative, bounded adaptation. In practice, it ensures that
the loop remains scalable by limiting actuation frequency and favoring stable primary control via $N_{\max}$.

Collectively, these design choices operationalize C-L4SaaS as an auditable, standards-aligned closed-loop workflow that links intent, enforcement, and NWDAF-driven assurance under bounded control and signaling constraints.


\section{ANALYTICAL MODEL AND CLOSED-LOOP CONTROL FORMULATION}

\subsection{System Model and Notation}

C-L4SaaS is modeled as a discrete-time closed-loop system sampled at the NWDAF analytics cadence. Let $n$ index the $n$-th monitoring window of duration $W$ seconds. The objective is to maintain compliance with service-level objectives (SLOs) for managed low-latency flows while operating under time-varying wireless capacity.

We consider a bottleneck resource in the end-to-end path (e.g., UPF egress, shared transport segment, or an equivalent constrained link). Let $q[n]$ be the bottleneck queue occupancy at the end of window $n$, $C[n]$ be the effective service capacity during window $n$ (time-varying in wireless), $\lambda[n]$ be the aggregate arrival rate during window $n$, and $\mu[n]\approx C[n]$ be the service rate. Let $T_s$ be the closed-loop update interval; in this paper, $T_s=W$.

A standard discrete-time queue evolution is
\begin{equation}
q[n+1]=\max\left\{0,\; q[n]+(\lambda[n]-\mu[n])T_s\right\}.
\end{equation}

The induced queuing delay component is approximated by
\begin{equation}
d_q[n]\approx \frac{q[n]}{C[n]}.
\end{equation}

This representation captures the primary challenge in 5G-Advanced environments: even when traffic demand is stable, volatility in $C[n]$ can create tail delay excursions unless policy adapts with appropriate guardrails.

\subsection{NWDAF Compliance Analytics as the Feedback Signal}

NWDAF produces compliance analytics from UPF measurements over each window $W$. C-L4SaaS focuses on outcome-based assurance; therefore, the feedback signal is expressed in terms of tail performance and loss rather than configuration state.

For a given managed low-latency flow (or an aggregate of a managed service set), define:
\begin{align}
\hat d_{99}(W_n) &= Q_{0.99}\big(\{d(t)\}_{t \in W_n}\big), \label{eq:tail-delay}\\
\hat \ell(W_n) &= \frac{N_{\mathrm{lost}}(W_n)}{N_{\mathrm{sent}}(W_n)}, \label{eq:loss-ratio}\\
\hat m(W_n) &= \frac{N_{\mathrm{CE}}(W_n)}{N_{\mathrm{ECNcap}}(W_n)}, \label{eq:ecn-ratio}\\
\hat r(W_n) &= \frac{B_{\mathrm{delivered}}(W_n)}{|W|}. \label{eq:delivered-throughput}
\end{align}

C-L4SaaS expresses compliance as predicates against SLO targets. Using 5QI~3 as a representative interactive low-latency class, the default targets are:
\begin{align}
\hat d_{99}(W_n) &\le d_{\mathrm{target}} = 50~\mathrm{ms}, \label{eq:delay-slo}\\
\hat \ell(W_n) &\le \ell_{\max} = 10^{-3}. \label{eq:loss-slo}
\end{align}

The feedback ``error'' is then defined as a windowed deviation from the tail-delay target:
\begin{equation}
e[n] = \hat d_{99}(W_n) - d_{\mathrm{target}}.
\label{eq:error}
\end{equation}

Optionally, a composite error can be defined if a joint delay--loss constraint is desired:
\begin{equation}
e_c[n] = w_d\big(\hat d_{99}(W_n)-d_{\mathrm{target}}\big) + w_\ell\big(\hat \ell(W_n)-\ell_{\max}\big),
\label{eq:composite-error}
\end{equation}
where $w_d$ and $w_\ell$ are weights selected by policy.

\subsection{Primary Actuation Variable and Bounded Update Law}

C-L4SaaS adopts a conservative actuation approach to ensure operational safety and stability. The primary PCF control variable is the admission limit $N_{\max}[n]$, which caps the number of concurrent managed low-latency sessions (or service grants) in a given policy domain.

The bounded update law is:
\begin{equation}
N_{\max}[n+1] = \Pi_{[0,\,N_{\mathrm{cap}}]}\bigl(N_{\max}[n] - K\,e[n]\bigr)
\end{equation}
where $K>0$ is the adaptation gain, $\Pi_{[0,\,N_{\mathrm{cap}}]}(\cdot)$ saturates the updated value into allowed bounds, and $N_{\mathrm{cap}}$ is a configured maximum derived from slice capacity planning and governance.

\textbf{Interpretation}. If the tail delay exceeds the target ($e[n]>0$), PCF reduces admission for subsequent grants by lowering $N_{\max}$. If the system remains compliant ($e[n]\le 0$) for a sustained period, PCF may gradually relax $N_{\max}$ back toward a planned operating point, subject to hysteresis and rate limits.

Secondary actions (priority offsets, GBR adjustments, enforcement profile switching) are treated as conditional, bounded measures used only when persistent non-compliance is detected, consistent with Table-I and Table-II in Section-II.

\subsection{Stability Condition and Anti-Oscillation Guardrails}

To reason about stability, C-L4SaaS uses a small-signal model around an operating point. Let the local sensitivity from admission control to tail-delay error be
\begin{equation}
G \triangleq \frac{\partial e}{\partial N_{\max}}.
\end{equation}
Assuming local linear behavior over one update step,
\begin{equation}
e[n{+}1] \approx e[n] - K G e[n] = (1 - K G) e[n].
\end{equation}
A sufficient condition for asymptotic stability of the linearized loop is
\begin{equation}
\lvert 1 - K G \rvert < 1
\;\Rightarrow\;
0 < K < \frac{2}{G}.
\end{equation}
Since $G$ depends on traffic mix and capacity volatility, C-L4SaaS enforces additional guardrails beyond the gain bound:
\begin{itemize}
  \item \textbf{Hysteresis margin:} treat ``Warning'' and ``Violating'' states with a margin $\delta$ (default 5~ms) to avoid reacting to noise.
  \item \textbf{Persistence:} require $k$ consecutive violating windows (default $k{=}3$) before stronger actions.
  \item \textbf{Rate limiting:} update $N_{\max}$ at most once per window $W$ and bound step sizes $\lvert\Delta N_{\max}\rvert$ to prevent abrupt oscillations.
  \item \textbf{Adaptive windowing:} if oscillation is detected, increase $W$ (e.g., from 2~s to 4~s) and freeze secondary actions until stability is restored.
\end{itemize}
These mechanisms are consistent with a practical operator control philosophy: prioritize stable service assurance over aggressive short-term corrections.

\subsection{Signaling Overhead Bounds for Closed-Loop Operation}

Closed-loop assurance consumes control-plane resources and must remain scalable. Let $N$ be the number of managed low-latency sessions under closed-loop monitoring, $B$ be the average message size per policy-update transaction (bytes), and $T_s$ be the update interval (seconds), with $T_s=W$. Then an upper bound on the average control traffic rate is
\begin{equation}
R_{\mathrm{ctrl}} \le \frac{N B}{T_s}.
\end{equation}
If the operator imposes a maximum allowable control overhead $R_{\max}$, the scalable population is bounded by
\begin{equation}
N \le \frac{R_{\max} T_s}{B}.
\end{equation}
This bound motivates admission control as the primary actuator: it allows the loop to protect tail delay while limiting the frequency and amplitude of control actions.

\subsection{Summary of Analytical Results}

This section formalized C-L4SaaS as a discrete-time feedback system in which NWDAF provides windowed compliance analytics and the PCF applies bounded policy adaptation, with admission control as the primary actuation mechanism. The stability condition $0 < K < 2/G$ provides a practical guideline for selecting conservative adaptation gains, while hysteresis, persistence, and rate limits mitigate oscillation under noisy measurements and volatile capacity. Finally, explicit overhead bounds quantify the scalability trade-off between responsiveness and control-plane load.

\section{DISCUSSION AND LIMITATIONS}

\subsection{Practical Deployment Considerations}


C-L4SaaS is implementable using existing 5GC functions, but requires strict governance of exposure, actuation, and assurance. NEF must enforce access control and rate limits, and intent-to-policy mapping must remain deterministic and auditable across subscriber, slice, and flow templates, including mid-session changes. Because UPF queue management and ECN marking are vendor-dependent, C-L4SaaS relies on policy-selected enforcement profiles and supports marking in either the RAN or UPF, provided NWDAF receives consistent compliance signals.

\subsection{Control Loop Safety and Responsiveness Trade-offs}


The framework prioritizes stability via bounded, rate-limited updates. Admission control \Nmax is the primary actuator to ensure predictable behavior under wireless volatility and to avoid oscillations from frequent low-level tuning; sudden capacity drops may still require tighter margins or bounded secondary actions (priority offsets or GBR within slice caps) to protect admitted sessions. The NWDAF window $W$ trades responsiveness for noise suppression: shorter windows react faster but can overreact to transients, while longer windows smooth noise but delay correction. Hysteresis and persistence mitigate oscillation, though parameter choices remain deployment-specific.

\subsection{Measurement Fidelity and Outcome Verification}

C-L4SaaS relies on NWDAF compliance analytics derived from UPF measurements. Delay estimation is sensitive to timestamp accuracy, sampling granularity, and aggregation policy. The framework therefore uses windowed tail metrics (e.g., $p99$ delay) and complements them with loss and ECN marking ratios to distinguish congestion-driven violations from measurement artifacts. Verified outcomes should be interpreted within the measurement scope: user-plane signals provide strong visibility into core and transport behavior, while radio scheduling effects may require additional context for end-to-end interpretation.

\subsection{Scope and Limitations}

This paper focuses on architecture and analytical formulation of closed-loop assurance for managed low-latency services. It provides stability and signaling overhead bounds and a standards-aligned control workflow, but does not claim empirical performance gains without deployment-specific validation. Future work will validate the framework under representative traffic mixes and variable-capacity conditions and quantify trade-offs between \Nmax-based admission control, bounded secondary actions, and analytics cadence across service classes and slice policies.

\subsection{Reproducible Evaluation Blueprint}

To support reproducibility, we outline an evaluation blueprint with three baselines: (i) no closed-loop assurance, (ii) L4S enablement without adaptation, and (iii) C-L4SaaS with NWDAF-driven adaptation. Representative scenarios include mixed L4S and classic traffic, time-varying wireless capacity, and controlled load surges. Metrics follow the NWDAF compliance set (tail delay, loss, CE-marking ratio, throughput) plus responsiveness indicators such as convergence time and policy update rate.

\section{Conclusion}
This paper presented C-L4SaaS, a closed-loop L4S-as-a-Service framework that advances managed low-latency service exposure and assurance in 5G-Advanced networks. By orchestrating NEF, PCF, NWDAF, and SMF/UPF, the framework links application intent to enforceable PCC rules and to outcome-based verification using NWDAF compliance analytics derived from user-plane measurements. Unlike static enablement approaches that stop at configuration, C-L4SaaS introduces bounded policy adaptation driven by verified service outcomes, with admission control as the primary actuator to preserve stability under volatile wireless conditions. The analytical formulation provides practical stability and signaling overhead bounds that guide safe parameter selection and scalable operation. 
Moreover, the governed exposure model enables differentiated low-latency service tiers and SLA-backed offerings grounded in verified outcomes, providing a pragmatic pathway for experience-driven monetization without relying on over-provisioning. Overall, C-L4SaaS establishes a standards-aligned foundation for exposing, assuring, and governing managed low-latency services, and provides a clear blueprint for subsequent implementation and validation in broader deployment contexts.

\end{document}